\documentclass[twocolumn]{bmcart}

\usepackage{amsthm,amsmath}
\RequirePackage{natbib}
\usepackage[utf8]{inputenc} 
\usepackage{comment}
\usepackage{graphicx}
\usepackage{hyperref}
\usepackage{float}
\usepackage{xcolor}
\usepackage{ifthen,color,comment,amssymb}
\usepackage{verbatim}
\usepackage{balance}
\usepackage{listings}

\usepackage{mathtools}

\DeclarePairedDelimiter\floor{\lfloor}{\rfloor}

\newboolean{showcomments}
\setboolean{showcomments}{true}
\ifthenelse{\boolean{showcomments}}
{ \newcommand{\mynote}[3]{
     \fbox{\bfseries\sffamily\scriptsize#1}
       {\small$\blacktriangleright$\textsf{\textcolor{#3}{{\em #2}\bf }}$\blacktriangleleft$}}}
       { \newcommand{\mynote}[2]{}}

\startlocaldefs
\endlocaldefs

\begin{document}

\begin{frontmatter}

\begin{fmbox}
\dochead{Research}


\title{Leveraging Blockchain for Immutable Logging and Querying Across Multiple Sites}


\author[
   addressref={aff1},                   
   corref={aff1},                       
   email={mustafa.ozdayi@utdallas.edu}   
]{\inits{MSO}\fnm{Mustafa Safa} \snm{Ozdayi}}
\author[
   addressref={aff1},
   email={muratk@utdallas.edu}
]{\inits{MK}\fnm{Murat} \snm{Kantarcioglu}}
\author[
   addressref={aff2},
   email={b.malin@vanderbilt.edu}
]{\inits{BM}\fnm{Bradley} \snm{Malin}}


\address[id=aff1]{
  \orgname{Department of Computer Science,
Erik Jonsson School of Engineering and Computer Science,
The University of Texas at Dallas,
800 W. Campbell Road,
Richardson, TX 75080, U.S.A.} 
}
\address[id=aff2]{%
   \orgname{Department of Computer Science, Vanderbilt University,
2301 Vanderbilt Place,
Nashville, TN 37235-1679,
U.S.A} 
}




\begin{abstractbox}

\begin{abstract}
\parttitle{Background} 
Blockchain has emerged as a decentralized and distributed framework that enables tamper-resilience and, thus, practical immutability for stored data. This immutability property is important in scenarios where auditability is desired, such as in maintaining access logs for sensitive healthcare and biomedical data. However, the underlying data structure of blockchain, by default, does not provide capabilities to efficiently query the stored data. In this investigation, we show that it is possible to efficiently run complex audit queries over the access log data stored on blockchains by using additional key-value stores. This paper specifically reports on the approach we designed for the blockchain track of iDASH Privacy \& Security Workshop 2018 competition. In this track, participants were asked to devise an efficient way to run conjunctive equality and range queries on a genomic dataset access log trail after storing it in a permissioned blockchain network consisting of 4 identical nodes, each representing a different site, created with the Multichain platform.
\parttitle{Methods}
Multichain duplicates and indexes blockchain data locally at each node in a key-value store to support retrieval requests at a later point in time. To efficiently leverage the key-value storage mechanism, we applied various techniques and optimizations, such as bucketization, simple data duplication and batch loading by accounting for the required query types of the competition and the interface provided by Multichain. Particularly, we implemented our solution and compared its loading and query-response performance with SQLite, a commonly used relational database, using the data provided by the iDASH 2018 organizers. 
\parttitle{Results}
Depending on the query type and the data size, the run time difference between blockchain based query-response and SQLite based query-response ranged from 0.2 seconds to 6 seconds. 
A deeper inspection revealed that range queries were the bottleneck of our solution which, nevertheless, scales up linearly.
\parttitle{Conclusions}
This investigation demonstrates that blockchain-based systems can provide reasonable query-response times to complex queries even if they only use simple key-value stores to manage their data. Consequently, we show that blockchains may be useful for maintaining data with auditability and immutability requirements across multiple sites. 


%
\end{abstract}




\end{abstractbox}
\end{fmbox}

\end{frontmatter}

\section*{Background}
\label{sec: Background}
Blockchains allow a set of parties to collaboratively maintain a collection of data on a tamper-resilient and decentralized ledger. This provides numerous benefits compared to traditional data storage models, where the administration of a shared database is delegated to either one or more trusted entities. 

One particularly notable benefit is that a blockchain-based data storage solution mitigates security issues that can arise from malicious administrators. Moreover, it eliminates the potential for a single point of failure because data is replicated across multiple entities. Due to their immutability and auditability guarantees, blockchains are useful for storing access logs from multiple sites to different datasets (e.g., genomic research data). This is because access logs require strong auditability (e.g., auditing accesses to genomic information), transparency (e.g., publicly verifying that certain data is not misused by checking the logs) and tamper-resistance (e.g., preventing an attacker from manipulating the stored logs). 

Furthermore, blockchains provide access to an uniform view of data that is logged from different sites. This, in conjunction with its other properties, can be beneficial in various contexts as demonstrated by previous investigations, such as those in the healthcare domain \cite{ozercan2018}\cite{choudhury2018}\cite{li2019dmms}.


However, the underlying data structure of a blockchain, by default, does not provide for an efficient technique to query the stored data. To overcome this limitation, most existing blockchain implementations, such as Bitcoin~\cite{Bitcoin} and Ethereum~\cite{ethereum}, provide support for key-value stores to duplicate and store data on the blockchain.
These key-value stores are then leveraged to support simple key-based retrieval queries that return the associated values. 

In this paper, we show how to leverage the key-value store provided by a blockchain to run complex queries efficiently (e.g., range queries) on the blockchain data. We specifically report on the approach we designed for the blockchain track of iDASH Privacy \& Security Workshop 2018 competition~\cite{iDash2018}. 
Our findings show  that  our  approach  induces
reasonable overhead, in terms query-response time, in
comparison to a traditional relational database management tool.

\subsection*{\textbf{Blockchain}}
Blockchain was first introduced by Nakamato as the underlying ledger of the now famous Bitcoin cryptocurrency~\cite{Bitcoin}. Briefly, a blockchain is an append-only, distributed and replicated database. It allows the participants of a network to collectively maintain a sequence of data in a tamper-resilient way. More importantly, it does so without a requirement for a trusted third party by invoking a consensus mechanism.

Informally, a blockchain network operates as follows: participants broadcast their data and certain nodes called \emph{miners} gather and store the data they receive in wrapper structures called \emph{blocks}. Through a consensus mechanism, the network elects a leader miner in a decentralized fashion for a sequence of epochs. The epoch leader broadcast his block to the network and, having received the leaders block, other nodes store it in their local memory where each block maintains a hash-link to the previous block. 

The consensus algorithm that the blockchain network deploys may depend on whether or not the network is \emph{permissionless}.
For example, Bitcoin operates on a permissionless network, where anyone is free to join and there is no uniform view of the network across participants. It utilizes a cryptographic puzzle called Proof-of-Work~\cite{PoW}  to achieve consensus. This makes tampering with the order of blocks computationally infeasible when the majority of the network participants follow the protocol honestly.

In \emph{permissioned} networks however, participants can employ more efficient consensus algorithms, such as PBFT~\cite{PBFT}. This is because the identity and number of participants are known to every party.

\subsection*{\textbf{Multichain}}
Multichain is a platform to deploy permissioned blockchains~\cite{Multichain}. In this context, permissioned means that access to the blockchain network can be arbitrarily restricted. Such networks are usually initialized by a single party who, at a later point in time, allocates permissions to other nodes to join the network and participate in the consensus protocol. For consensus, Multichain deploys a variant of a classical Byzantine fault tolerance algorithm whose exact details are provided in the \emph{Mining in MultiChain} section of the corresponding whitepaper \cite{MultiChainWhite}.

To handle queries efficiently, Multichain provides a module called \emph{streams}, which uses an abstraction of a dictionary (i.e., key-value store) on top of a blockchain~\cite{Streams}. The streams module allows a node to store an arbitrary datum and an associated key by submitting a key-value pair in a transaction to the blockchain. Multichain duplicates and indexes the data stored on the blockchain in LevelDB (a key-value store~\cite{LevelDB}), which is locally maintained by each node to serve queries submitted to the blockchain efficiently~\cite{MultichainStreams}. In other words, the streams module allows a node to interact with the underlying key-value store.
It is possible to store multiple values with the same key, such that query results can be returned as lists. 

The streams module supports
the following methods (among others) on top of a blockchain.
\begin{itemize}
    \item \textbf{createDictionary}(dictionary-name): Creates a dictionary with the specified name.
    \item \textbf{insert}(dictionary-name, value, key): Inserts the key-value pair to the specified dictionary.
    \item \textbf{retrieve}(dictionary-name, key): Retrieves the value(s) corresponding to the given key from the specified dictionary.
\end{itemize}
 We note that it is possible to create an arbitrary number of dictionaries on top of a blockchain, each of which stores data independently.
 
\subsection*{\textbf{Overview of the task}}
The blockchain track of the iDASH Privacy \& Security Workshop 2018 competition provided a genomic dataset access log trail~\cite{iDASH2018Tracks} in which each log consists of 7 fields:
\emph{timestamp, node, ID, ref-ID, user, activity} and \emph{resource}. The activity and resource fields can take on arbitrary string values. The other fields can take on arbitrary positive integer values.

This trail is stored on a blockchain, where it is assumed that the trail arrives as a data stream (i.e., one log at a time). The competition rules dictated that the blockchain must be created using Multichain version 1.0.4 and the network must consist of four identical nodes, each representing a different site, initialized with default parameters per Multichain specifications. It should be possible to insert and query the data using any node.
Also, the rules prohibited the use of using any off-chain mechanisms to handle the data other than what Multichain provides.

The goal is to develop a system that can query the blockchain efficiently under this setting while minimizing loading times and storage space.
A viable solution must support two types of queries:
\begin{itemize}
    \item Conjunctive equality queries on selected fields.
    \item A range query on the timestamp field.
\end{itemize}
Furthermore, the system should support the reporting of results in ascending or descending order on any field.

\section*{Methods}
\label{sec: Our_Sol}
We now describe the techniques and the optimizations we deployed to handle queries efficiently in our system.
We note that although our system is explicitly tuned for the blockchain track of iDASH 2018, our methods can be applied to support more general queries.

In what follows, we first describe how to handle conjunctive equality queries. Next, we describe how to handle range queries. We then explain how query response times can be further improved via batch loading.

\subsection*{\textbf{Conjunctive Equality Queries}}
For each field, we create a dictionary that uses field values as keys and logs as values. For example, in the \emph{user} dictionary at key $1$, we have the logs whose user field's value is $1$. Similarly, in the \emph{ID} dictionary at key $2$, we have the logs whose ID is $2$. Note that the logs are duplicated for each field. 

When processing such a query, we first find the most restrictive field key and retrieve the logs from the corresponding dictionary with that key. Next, we filter the retrieved logs with other field keys. Finding the most restrictive field can be achieved efficiently. This is because Multichain keeps track of how many items are stored at a key in a dictionary, which can be accessed by a \textbf{getCount}(dictionary-name, key) method.

As an example, consider a query that requests logs with $user=1 \wedge ID=2$. We first compute $x$ = \textbf{getCount}(user-dictionary, 1) and $y$ = \textbf{getCount}(id-dictionary, 2).
Then if $x > y$, we retrieve the logs from the id-dictionary, via \textbf{retrieve}(id-dictionary, 2) and discard the logs whose user field is not equal to 1.

\subsection*{\textbf{Range Queries on a Single Field}}
To handle range queries, we designed a \emph{bucketization} technique.
That is, we create intervals of a fixed size and assign each log to exactly one of those intervals depending on the queried field's value. Each interval is referred as a \emph{bucket} and identified by an unique value. Particularly, our bucketization technique works as follows: first, we create a separate dictionary, \emph{range-dictionary}, in which we assign each log \emph{key} $= \floor{\text{Timestamp of log}/N}$, where $N$ is a predefined bucket size.

Next, given a range query $[x, y]$, which requests logs whose timestamps are between $x$ and $y$ (inclusive), we retrieve all of the logs with keys $\floor{x/N} + 1, \floor{x/N} + 2, \dots, \floor{y/N}-1$. 
Finally, we retrieve and perform a linear scan of the logs at keys $\floor{x/N}$ and $\floor{y/N}$ and discard logs whose timestamps are not in $[x, y]$.

Note that, for each log with a key in range $[\floor{x/N} + 1, \floor{y/N}-1]$, it is guaranteed that the log
is in $[x, y]$. As a result, we do not need to scan the logs stored at these keys. 

\subsection*{\textbf{Improving Retrieval Speed via Batch Loading}}
We observed that in the Multichain platform loading logs in batches can substantially improve retrieval speed. 
Here batch loading means that, instead of inserting one log in each transaction, we buffer and insert several logs in a single transaction.

We observed that if we load logs as batches of size $k$, then retrieving these logs would be roughly $k$ times faster than storing them one at a time. The competition rules required the solution to be crash-resistant, so a straightforward way of buffering would have failed to meet this requirement. For example, if our buffer size is $4$, then we load logs to the blockchain in batches of size 4. Yet if the system crashes after the first 2 logs arrive, then both of these logs would be lost due to the fact that the contents of the buffer was not loaded to the blockchain at the crash time.

To overcome this problem, we extend our solution to maintain two dictionaries per field, a \emph{batch dictionary} and a \emph{regular dictionary}. In the batch dictionary, we load logs in batches, while in a regular dictionary, we load the logs one at a time  (i.e., a buffer size of 1). When retrieving data, we select from the batch dictionary, compare the size of the retrieved list with the corresponding list in the regular dictionary and execute a crash recovery (if needed).

For example, imagine a query that attempts to retrieve logs with $node=1$. To support this query, we first retrieve the logs from the batch dictionary via \emph{batchLogs} = \textbf{retrieve}(batch-node-dictionary, 1). We then compute the length of the corresponding list in the regular dictionary via \emph{logListSize} = \textbf{getCount}(regular-node-dictionary, 1). Then, if the size of \emph{batchLogs} is equal to \emph{logListSize}, we simply return \emph{batchLogs} as the result. 

Otherwise, it becomes evident
that we lost some data from the batch dictionary due to a crash. To recover, we compute the difference between the size of \emph{batchLogs} and \emph{logListSize},
i.e., $x$ = \emph{logListSize} - \textbf{size}(batchLogs). We then retrieve the last $x$ items from the regular dictionary and append these to both \emph{batchLogs} and the batch dictionary. Finally, we return \emph{batchLogs} as the result.

\section*{Results}
\label{sec:Experiments}
In this section, we report on a set of  experiments designed to characterize the performance of our system. 

\subsection*{\textbf{Implementation details and experiment setting}}

We implemented our solution using Python 3.5.2, Multichain 1.0.4 and used the Savoir wrapper to interact with Multichain API~\cite{Savoir}. Our code is available at~\cite{MyEntry}. Our test setup consisted of four identical virtual machines with the following specifications: 2-Core CPU (2.6. GHz Intel Xeon E5), 7.5 GB of RAM and 50 GB of storage with the Ubuntu 14.04 LTS operating system. 

We used the dataset supplied by the competition organizers, which consisted of four files, one per node, in which each file has $10^5$ logs. The following is an illustration of the structure and content of the logs:
{\small 
\begin{verbatim}
TIMESTAMP	NODE ID REF-ID USER ACTIVITY RESOURCE
1522000002801	1	1	1	1	REQ_RESOURCE	TOPMed
1522000008352	1	2	1	1	VIEW_RESOURCE	GTEx
1522000016966	1	3	3	6	REQ_RESOURCE	MOD_FlyBase
1522000019451	1	4	1	1	FILE_ACCESS TOPMed
1522000019566	1	5	3	6	VIEW_RESOURCE	MOD_FlyBase
1522000024848	1	6	6	10	FILE_ACCESS GTEx
\end{verbatim}}
 \vspace{0.2cm}
 
From the previous discussion regarding batch loading, it is evident that the larger the buffer size, the faster the retrieval speed. However, Multichain imposes a size limit on each transaction, such that it is not possible to increase buffer size arbitrarily. We observed that, for the given dataset, the transaction size limit is reached for a buffer size around $10^4$. As a result, we set the buffer size accordingly.

Now, it can be seen that the number of buckets we have to retrieve decreases with the increasing bucket size. However, the number of individual logs we have to scan may increase. This is because it depends on the distribution of logs over the buckets and the query. Note that if one chooses bucket size poorly, it could be the case that all the logs would go to the same bucket.

To determine an appropriate bucket size, we ran several range queries of varying sizes on the given data and measured the average running time. During our empirical analysis, we observed an increase in the average running time as bucket size increased from $10^1$ to $10^7$. After that point however, average running time started to decrease with the increasing bucket size. As a result, we selected a bucket size of $10^7.$

Finally, we compared our solution's performance with a traditional relational database, namely SQLite 3.22 which we ran in one of the virtual machines. We report on the average measures over 10 runs and illustrate standard deviations by error bars in our plots.

\subsection*{\textbf{Experiments}}
\begin{figure}[]
\centering
\includegraphics[scale=0.28]{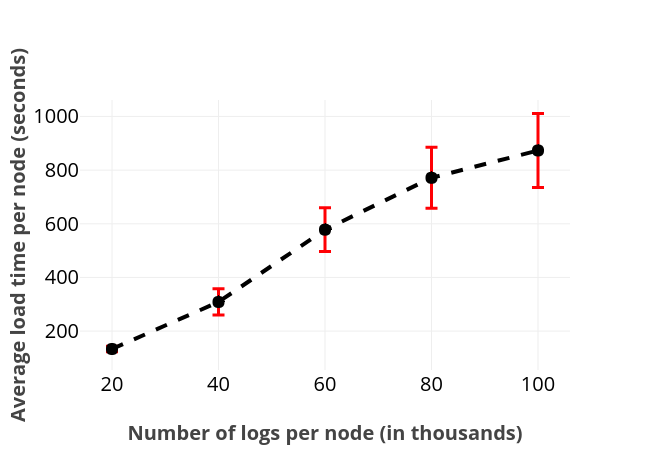}
\caption{Average time required for a node to load logs of various sizes. The large standard deviation is likely due to network latency. As expected, load times scale linearly with the number of the logs.}
\label{fig:Loading_logs}
\end{figure}

\begin{figure}[]
\centering
\includegraphics[scale=0.28]{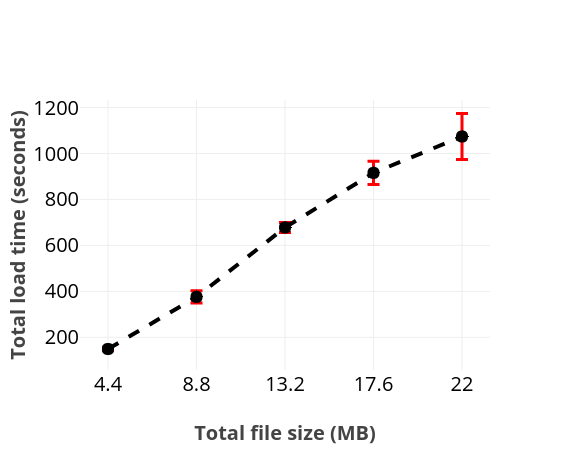}
\caption{Total load time required as a function of file size.  As expected, the total load time scaled linearly in the size of the file.}
\label{fig:Loading_files}
\end{figure}

First, we measured the load time by using logs of various sizes concurrently at each node. Figure~\ref{fig:Loading_logs} depicts the average load time of a node with respect to number of logs loaded by it. We further plot  the influence of file size on total load time, which corresponds to the slowest node. Those results are provided in Figure~\ref{fig:Loading_files}. These figures do not include the results from SQLite because our solution is substantially slower. For instance, loading all $400.000$ logs required merely about 4 seconds in SQLite.

Next, we investigated query-response times. We ran the test queries supplied by the competition organizers. The queries and the number of records returned by were as follows.

\begin{figure}[]
\centering
\includegraphics[scale=0.28]{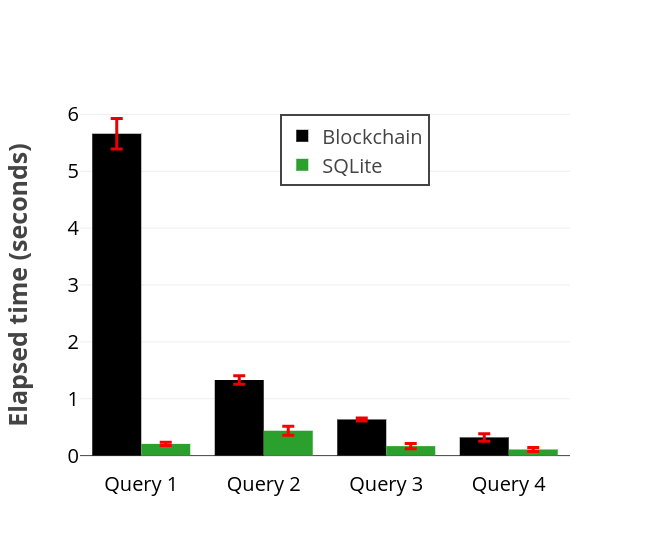}
\caption{Running times of test queries. Query 1 is a range query and the others are conjunctive equality queries. Results imply range queries dominate the performance.}
\label{fig:Test_queries}
\end{figure} 

\newpage
\begin{itemize}

\item{\begin{lstlisting}[language=SQL]
Query 1: SELECT * FROM Chain WHERE 
user = 7 AND (timestamp BETWEEN 
1522257730000 AND 1522449160000). 
Returns 30489 records.
\end{lstlisting}}
\item{\begin{lstlisting}[language=SQL]
Query 2: SELECT * FROM Chain WHERE 
resource = 'MOD_WormBase'. 
Returns 67462 records.
\end{lstlisting}}
\item{\begin{lstlisting}[language=SQL] 
Query 3: SELECT * FROM Chain WHERE 
user = 1 AND resource = 'TOPMed'. 
Returns 17098 records.
\end{lstlisting}}
\item{\begin{lstlisting}[language=SQL]
Query 4: SELECT * FROM Chain WHERE 
node = 3 AND ref-ID = 40345 
ORDER BY timestamp ASC. 
Returns 5983 records.
\end{lstlisting}}

\end{itemize}

The running times for these queries are shown Figure~\ref{fig:Test_queries} As the results indicate, the main bottleneck of our solution is the range query. We also note that SQLite's internal representation and processing scheme is quite different than our method. As such, the SQLite running time is not always highly correlated with the blockchain time.

\begin{figure}[]
\centering
\includegraphics[scale=0.25]{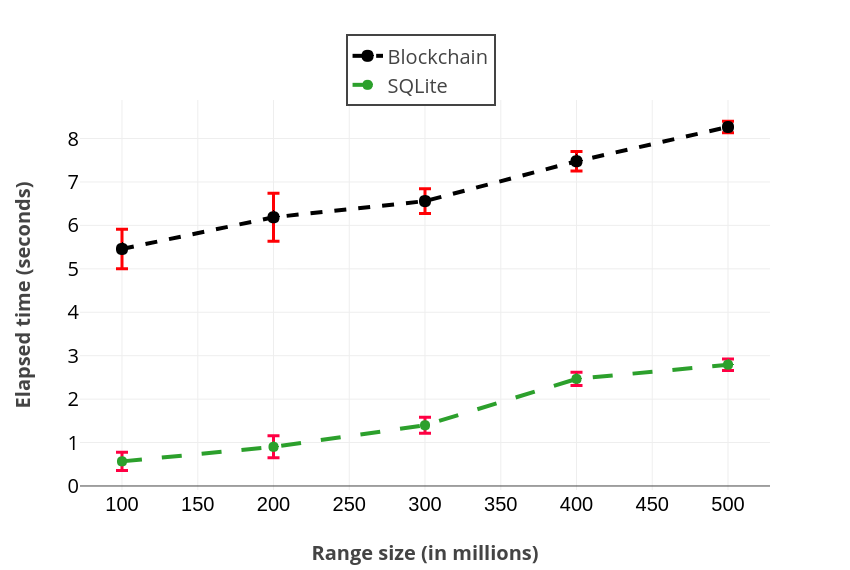}
\caption{Running times of range queries.}
\label{fig:Range_queries}
\end{figure}

In Figure~\ref{fig:Range_queries}, we compare the range query performance of our solution with respect to SQLite's performance.  We observe both methods scale linearly where the difference is between 5-6 seconds.

\begin{figure}[]
\centering
\includegraphics[scale=0.25]{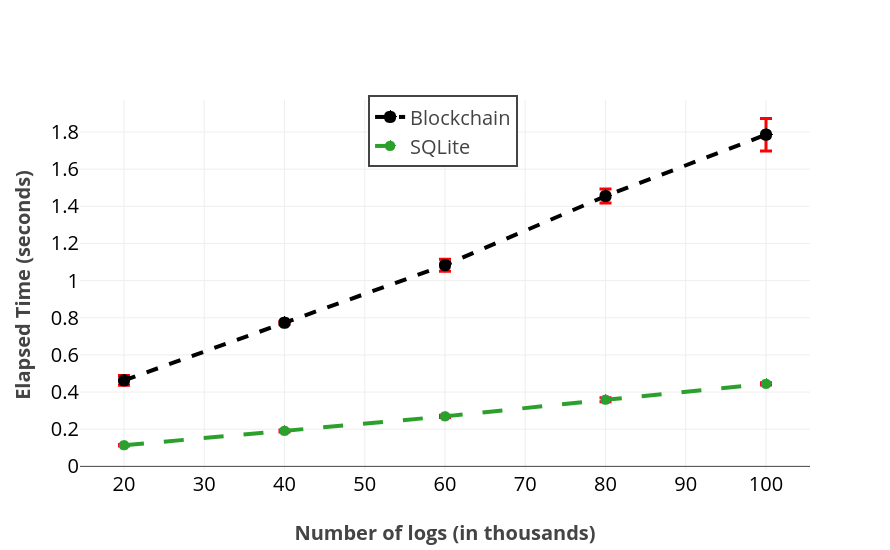}
\caption{Processing time required as a function of the number of retrieved records.}
\label{fig:Retrieval_times}
\end{figure}

Finally, Figure~\ref{fig:Retrieval_times} illustrates how the number of records retrieved influences the query-response time. In this experiment, we ran queries without any restrictions (i.e., query returns every stored log) after loading appropriate number of logs.
Given how our approach handles conjunctive equality queries, this plot also represents the performance of conjunctive equality queries. This is due to the fact that a conjunctive equality query makes a call to \textbf{getCount}(.) per field given in the query in addition to retrieving the data. This only adds a negligible overhead.

Finally, we considered the storage requirements. After loading all logs, the size of the blockchain was about 3 GBs per node, whereas the size of the SQLite database was several orders of magnitude smaller at 20 MBs.

\section*{Discussion}
In this section, we discuss certain limitations and highlight opportunities for improvement of our approach.

First, as mentioned earlier, bucket and buffer size was based on an empirical investigation. We did not conduct extensive studies on these parameters to optimize them. It might be possible to improve query-response times by fine-tuning these parameters.

Second, it is possible to map string fields (i.e., resource and activity) to the integer values to reduce the size of logs. This may improve the loading and query-response times.

Third, we did not consider parallelization. Although Multichain platform imposes some limitations on the parallelization (e.g., concurrently reading different parts of a stream is not possible) workarounds might exist~\cite{Parallelization}.

Further, per the rules of the competition, we were not permitted to modify the blockchain parameters. A straightforward way of improving performance might be to optimize these parameters. For example, the \emph{target-block-time} parameter controls the average number of seconds between two blocks whose default value is 15. It might be possible to decrease loading times by letting the blockchain generates blocks more often.

Finally, we note that Multichain is expected to deploy some new features to support data handling more efficiently in future versions. For instance, in version 2, blockchain stores just the hashes of data~\cite{Multichain2}. Since transactions will be shortened, this will likely reduce loading and response times. One can simply compare the hashes of data after fetching them from the accompanying key-value store with the hash on blockchain to ensure immutability in this model.

\section*{Conclusion}
\label{sec:Conclusion}
In this paper, we demonstrated that blockchain technology can overcome inherent limitations on querying and, thus, can be a useful tool for managing data accross multiple sites, particularly in scenarios that require strong immutability and auditability. We showed how bucketization, simple data duplication and batch loading can be utilized to run complex complex queries efficiently over blockchains that provide support for only simple key-value stores. Particularly, we implemented these notions in the submission to the blockchain track of iDASH 2018 competition that supports efficient conjunctive equality and range queries over blockchains created with Multichain platform. We illustrated that our approach induced reasonable overhead, in terms query-response time, in comparison to a traditional relational database management tool.

\balance


\begin{backmatter}


\subsection*{\textbf{Abbreviations}}
SQL: Structured Query Language; PBFT: Practical Byzantine Fault Tolerance.

\subsection*{\textbf{Ethics approval and consent to participate}}
Not applicable.

\subsection*{\textbf{Consent for publication}}
Not applicable.

\subsection*{\textbf{Availability of data and material}}

Dataset is available at~\cite{iDASH2018Tracks}
and our implementation is available at~\cite{MyEntry}.

\subsection*{\textbf{Competing interests}}
The authors declare that they have no competing interests.

\subsection*{\textbf{Acknowledgments}}
The authors would like to thank the anonymous reviewers for their constructive suggestions and comments.

\subsection*{\textbf{Funding}}
This publication was partly supported by NIH award 1R01HG006844, NSF awards CICI- 1547324, IIS-1633331, CNS-1837627, OAC-1828467 and ARO award W911NF-17-1-0356.

\subsection*{\textbf{Author's contributions}}
All authors contributed equally to this work. All authors have read and approved this manuscript.



\bibliographystyle{bmc-mathphys} 
\bibliography{bmc_article}      


\begin{thebibliography}{18}
\ifx \bisbn   \undefined \def \bisbn  #1{ISBN #1}\fi
\ifx \binits  \undefined \def \binits#1{#1}\fi
\ifx \bauthor  \undefined \def \bauthor#1{#1}\fi
\ifx \batitle  \undefined \def \batitle#1{#1}\fi
\ifx \bjtitle  \undefined \def \bjtitle#1{#1}\fi
\ifx \bvolume  \undefined \def \bvolume#1{\textbf{#1}}\fi
\ifx \byear  \undefined \def \byear#1{#1}\fi
\ifx \bissue  \undefined \def \bissue#1{#1}\fi
\ifx \bfpage  \undefined \def \bfpage#1{#1}\fi
\ifx \blpage  \undefined \def \blpage #1{#1}\fi
\ifx \burl  \undefined \def \burl#1{\textsf{#1}}\fi
\ifx \doiurl  \undefined \def \doiurl#1{\textsf{#1}}\fi
\ifx \betal  \undefined \def \betal{\textit{et al.}}\fi
\ifx \binstitute  \undefined \def \binstitute#1{#1}\fi
\ifx \binstitutionaled  \undefined \def \binstitutionaled#1{#1}\fi
\ifx \bctitle  \undefined \def \bctitle#1{#1}\fi
\ifx \beditor  \undefined \def \beditor#1{#1}\fi
\ifx \bpublisher  \undefined \def \bpublisher#1{#1}\fi
\ifx \bbtitle  \undefined \def \bbtitle#1{#1}\fi
\ifx \bedition  \undefined \def \bedition#1{#1}\fi
\ifx \bseriesno  \undefined \def \bseriesno#1{#1}\fi
\ifx \blocation  \undefined \def \blocation#1{#1}\fi
\ifx \bsertitle  \undefined \def \bsertitle#1{#1}\fi
\ifx \bsnm \undefined \def \bsnm#1{#1}\fi
\ifx \bsuffix \undefined \def \bsuffix#1{#1}\fi
\ifx \bparticle \undefined \def \bparticle#1{#1}\fi
\ifx \barticle \undefined \def \barticle#1{#1}\fi
\ifx \bconfdate \undefined \def \bconfdate #1{#1}\fi
\ifx \botherref \undefined \def \botherref #1{#1}\fi
\ifx \url \undefined \def \url#1{\textsf{#1}}\fi
\ifx \bchapter \undefined \def \bchapter#1{#1}\fi
\ifx \bbook \undefined \def \bbook#1{#1}\fi
\ifx \bcomment \undefined \def \bcomment#1{#1}\fi
\ifx \oauthor \undefined \def \oauthor#1{#1}\fi
\ifx \citeauthoryear \undefined \def \citeauthoryear#1{#1}\fi
\ifx \endbibitem  \undefined \def \endbibitem {}\fi
\ifx \bconflocation  \undefined \def \bconflocation#1{#1}\fi
\ifx \arxivurl  \undefined \def \arxivurl#1{\textsf{#1}}\fi
\csname PreBibitemsHook\endcsname

\bibitem{ozercan2018}
\begin{barticle}
\bauthor{\bsnm{Ozercan}, \binits{H.I.}},
\bauthor{\bsnm{Ileri}, \binits{A.M.}},
\bauthor{\bsnm{Ayday}, \binits{E.}},
\bauthor{\bsnm{Alkan}, \binits{C.}}:
\batitle{Realizing the potential of blockchain technologies in genomics}.
\bjtitle{Genome research}
\bvolume{28}(\bissue{9}),
\bfpage{1255}--\blpage{1263}
(\byear{2018})
\end{barticle}
\endbibitem

\bibitem{choudhury2018}
\begin{botherref}
\oauthor{\bsnm{Choudhury}, \binits{O.}},
\oauthor{\bsnm{Sarker}, \binits{H.}},
\oauthor{\bsnm{Rudolph}, \binits{N.}},
\oauthor{\bsnm{Foreman}, \binits{M.}},
\oauthor{\bsnm{Fay}, \binits{N.}},
\oauthor{\bsnm{Dhuliawala}, \binits{M.}},
\oauthor{\bsnm{Sylla}, \binits{I.}},
\oauthor{\bsnm{Fairoza}, \binits{N.}},
\oauthor{\bsnm{Das}, \binits{A.K.}}:
Enforcing human subject regulations using blockchain and smart contracts.
Blockchain in Healthcare Today
(2018)
\end{botherref}
\endbibitem

\bibitem{li2019dmms}
\begin{botherref}
\oauthor{\bsnm{Li}, \binits{P.}},
\oauthor{\bsnm{Nelson}, \binits{S.D.}},
\oauthor{\bsnm{Malin}, \binits{B.A.}},
\oauthor{\bsnm{Chen}, \binits{Y.}}:
Dmms: A decentralized blockchain ledger for the management of medication
  histories.
Blockchain in Healthcare Today
(2019)
\end{botherref}
\endbibitem

\bibitem{Bitcoin}
\begin{botherref}
\oauthor{\bsnm{Nakamoto}, \binits{S.}}:
Bitcoin: A peer-to-peer electronic cash system
(2008)
\end{botherref}
\endbibitem

\bibitem{ethereum}
\begin{barticle}
\bauthor{\bsnm{Wood}, \binits{G.}}:
\batitle{Ethereum: A secure decentralised generalised transaction ledger}.
\bjtitle{Ethereum project yellow paper}
\bvolume{151},
\bfpage{1}--\blpage{32}
(\byear{2014})
\end{barticle}
\endbibitem

\bibitem{iDash2018}
\begin{botherref}
iDASH Secure Genome Analysis Competition 2018, GMC Medical Genomics, 2019
\end{botherref}
\endbibitem

\bibitem{PoW}
\begin{bbook}
\bauthor{\bsnm{Jakobsson}, \binits{M.}},
\bauthor{\bsnm{Juels}, \binits{A.}}:
In: \beditor{\bsnm{Preneel}, \binits{B.}} (ed.)
\bbtitle{Proofs of Work and Bread Pudding Protocols(Extended Abstract)},
pp. \bfpage{258}--\blpage{272}.
\bpublisher{Springer},
\blocation{Boston, MA}
(\byear{1999})
\end{bbook}
\endbibitem

\bibitem{PBFT}
\begin{barticle}
\bauthor{\bsnm{Castro}, \binits{M.}},
\bauthor{\bsnm{Liskov}, \binits{B.}}:
\batitle{Practical byzantine fault tolerance and proactive recovery}.
\bjtitle{ACM Trans. Comput. Syst.}
\bvolume{20}(\bissue{4}),
\bfpage{398}--\blpage{461}
(\byear{2002}).
doi:\doiurl{10.1145/571637.571640}
\end{barticle}
\endbibitem

\bibitem{Multichain}
\begin{botherref}
Multichain: An Open Platform for Building Blockchains.
\url{https://www.multichain.com/}
Accessed 2 June 2019
\end{botherref}
\endbibitem

\bibitem{MultiChainWhite}
\begin{botherref}
MultiChain Private Blockchain — White Paper.
\url{https://www.multichain.com/download/MultiChain-White-Paper.pdf}
Accessed 2 June 2019
\end{botherref}
\endbibitem

\bibitem{Streams}
\begin{botherref}
Introducing MultiChain Streams.
\url{https://www.multichain.com/blog/2016/09/introducing-multichain-streams/}
Accessed 2 June 2019
\end{botherref}
\endbibitem

\bibitem{LevelDB}
\begin{botherref}
LevelDB.
\url{http://leveldb.org/}
Accessed 2 June 2019
\end{botherref}
\endbibitem

\bibitem{MultichainStreams}
\begin{botherref}
How do Streams Work Under the Hood?
\url{https://www.multichain.com/qa/9635/how-do-streams-work-under-the-hood}
Accessed 2 June 2019
\end{botherref}
\endbibitem

\bibitem{iDASH2018Tracks}
\begin{botherref}
iDASH Privacy and Security Workshop 2018 Competition Tracks.
\url{www.humangenomeprivacy.org/2018/competition-tasks.html}
Accessed 2 June 2019
\end{botherref}
\endbibitem

\bibitem{Savoir}
\begin{botherref}
Savoir:A Python Wrapper for Multichain Json-RPC API.
\url{https://github.com/DXMarkets/Savoir}
Accessed 2 June 2019
\end{botherref}
\endbibitem

\bibitem{MyEntry}
\begin{botherref}
\url{https://github.com/TinfoilHat0/idash2018BlockchainTrack}
Accessed 2 June 2019
\end{botherref}
\endbibitem

\bibitem{Parallelization}
\begin{botherref}
Multiprocessing and Streams.
\url{https://www.multichain.com/qa/10947/multiprocessing-and-streams}
Accessed 2 June 2019
\end{botherref}
\endbibitem

\bibitem{Multichain2}
\begin{botherref}
Second MultiChain 2.0 Preview Release.
\url{https://www.multichain.com/blog/2018/01/second-multichain-2-0-preview-release/}
Accessed 2 June 2019
\end{botherref}
\endbibitem

\end{thebibliography}

\newcommand{\BMCxmlcomment}[1]{}

\BMCxmlcomment{

<refgrp>

<bibl id="B1">
  <title><p>Realizing the potential of blockchain technologies in
  genomics</p></title>
  <aug>
    <au><snm>Ozercan</snm><fnm>HI</fnm></au>
    <au><snm>Ileri</snm><fnm>AM</fnm></au>
    <au><snm>Ayday</snm><fnm>E</fnm></au>
    <au><snm>Alkan</snm><fnm>C</fnm></au>
  </aug>
  <source>Genome research</source>
  <publisher>Cold Spring Harbor Lab</publisher>
  <pubdate>2018</pubdate>
  <volume>28</volume>
  <issue>9</issue>
  <fpage>1255</fpage>
  <lpage>-1263</lpage>
</bibl>

<bibl id="B2">
  <title><p>Enforcing human subject regulations using blockchain and smart
  contracts</p></title>
  <aug>
    <au><snm>Choudhury</snm><fnm>O</fnm></au>
    <au><snm>Sarker</snm><fnm>H</fnm></au>
    <au><snm>Rudolph</snm><fnm>N</fnm></au>
    <au><snm>Foreman</snm><fnm>M</fnm></au>
    <au><snm>Fay</snm><fnm>N</fnm></au>
    <au><snm>Dhuliawala</snm><fnm>M</fnm></au>
    <au><snm>Sylla</snm><fnm>I</fnm></au>
    <au><snm>Fairoza</snm><fnm>N</fnm></au>
    <au><snm>Das</snm><fnm>AK</fnm></au>
  </aug>
  <source>Blockchain in Healthcare Today</source>
  <pubdate>2018</pubdate>
</bibl>

<bibl id="B3">
  <title><p>DMMS: A Decentralized Blockchain Ledger for the Management of
  Medication Histories</p></title>
  <aug>
    <au><snm>Li</snm><fnm>P</fnm></au>
    <au><snm>Nelson</snm><fnm>SD</fnm></au>
    <au><snm>Malin</snm><fnm>BA</fnm></au>
    <au><snm>Chen</snm><fnm>Y</fnm></au>
  </aug>
  <source>Blockchain in Healthcare Today</source>
  <pubdate>2019</pubdate>
</bibl>

<bibl id="B4">
  <title><p>Bitcoin: A peer-to-peer electronic cash system</p></title>
  <aug>
    <au><snm>Nakamoto</snm><fnm>S</fnm></au>
  </aug>
  <pubdate>2008</pubdate>
</bibl>

<bibl id="B5">
  <title><p>Ethereum: A secure decentralised generalised transaction
  ledger</p></title>
  <aug>
    <au><snm>Wood</snm><fnm>G</fnm></au>
  </aug>
  <source>Ethereum project yellow paper</source>
  <pubdate>2014</pubdate>
  <volume>151</volume>
  <fpage>1</fpage>
  <lpage>-32</lpage>
</bibl>

<bibl id="B6">
  <title><p>iDASH secure genome analysis competition 2018, GMC Medical
  Genomics, 2019</p></title>
</bibl>

<bibl id="B7">
  <title><p>Proofs of Work and Bread Pudding Protocols(Extended
  Abstract)</p></title>
  <aug>
    <au><snm>Jakobsson</snm><fnm>M</fnm></au>
    <au><snm>Juels</snm><fnm>A</fnm></au>
  </aug>
  <source>Secure Information Networks: Communications and Multimedia Security
  IFIP TC6/TC11 Joint Working Conference on Communications and Multimedia
  Security (CMS'99) September 20--21, 1999, Leuven, Belgium</source>
  <publisher>Boston, MA: Springer US</publisher>
  <editor>Preneel, Bart</editor>
  <pubdate>1999</pubdate>
  <fpage>258</fpage>
  <lpage>-272</lpage>
</bibl>

<bibl id="B8">
  <title><p>Practical Byzantine Fault Tolerance and Proactive
  Recovery</p></title>
  <aug>
    <au><snm>Castro</snm><fnm>M</fnm></au>
    <au><snm>Liskov</snm><fnm>B</fnm></au>
  </aug>
  <source>ACM Trans. Comput. Syst.</source>
  <publisher>New York, NY, USA: ACM</publisher>
  <pubdate>2002</pubdate>
  <volume>20</volume>
  <issue>4</issue>
  <fpage>398</fpage>
  <lpage>-461</lpage>
  <url>http://doi.acm.org/10.1145/571637.571640</url>
</bibl>

<bibl id="B9">
  <title><p>Multichain: An open platform for building blockchains</p></title>
  <url>https://www.multichain.com/</url>
</bibl>

<bibl id="B10">
  <title><p>MultiChain Private Blockchain — White Paper</p></title>
  <url>https://www.multichain.com/download/MultiChain-White-Paper.pdf</url>
</bibl>

<bibl id="B11">
  <title><p>Introducing MultiChain Streams</p></title>
  <url>https://www.multichain.com/blog/2016/09/introducing-multichain-streams/</url>
</bibl>

<bibl id="B12">
  <title><p>LevelDB</p></title>
  <url>http://leveldb.org/</url>
</bibl>

<bibl id="B13">
  <title><p>How do streams work under the hood?</p></title>
  <url>https://www.multichain.com/qa/9635/how-do-streams-work-under-the-hood</url>
</bibl>

<bibl id="B14">
  <title><p>iDASH Privacy and Security Workshop 2018 Competition
  Tracks</p></title>
  <url>www.humangenomeprivacy.org/2018/competition-tasks.html</url>
</bibl>

<bibl id="B15">
  <title><p>Savoir:A python wrapper for Multichain Json-RPC API</p></title>
  <url>https://github.com/DXMarkets/Savoir</url>
</bibl>

<bibl id="B16">
  <url>https://github.com/TinfoilHat0/idash2018BlockchainTrack</url>
</bibl>

<bibl id="B17">
  <title><p>Multiprocessing and streams</p></title>
  <url>https://www.multichain.com/qa/10947/multiprocessing-and-streams</url>
</bibl>

<bibl id="B18">
  <title><p>Second MultiChain 2.0 preview release</p></title>
  <url>https://www.multichain.com/blog/2018/01/second-multichain-2-0-preview-release/</url>
</bibl>

</refgrp>
} 






\end{backmatter}
\end{document}